\documentclass[fleqn,usenatbib]{mnras}

\usepackage[T1]{fontenc}
\usepackage{ae,aecompl}
\usepackage{subfig}
\usepackage{graphicx}	
\usepackage{amsmath}	
\usepackage{amssymb}	
\usepackage{color}
\usepackage{xcolor}
\usepackage{mathptmx}   
\usepackage{booktabs}
\usepackage{lipsum}
\usepackage{widetext}
\usepackage{rotating}
\usepackage{pdflscape}

\title[Galaxy Cluster Mis-centring]{An analysis of galaxy cluster mis-centring using cosmological hydrodynamic simulations}

\author[Z. Yan et al.]{
Z. Yan, $^{1}$\thanks{E-mail:
yanza15@phas.ubc.ca}
N. Raza,$^{1}$\thanks{E-mail:
nraza@phas.ubc.ca}
L. Van Waerbeke,$^{1}$\thanks{E-mail:
waerbeke@phas.ubc.ca}
A. J. Mead,$^{1,2}$
I. G. McCarthy,$^{3}$
\newauthor
T. Tr\"{o}ster$^{4}$
and G. Hinshaw$^{1}$
\\
$^{1}$Department of Physics and Astronomy, University of British Columbia, 6224 Agricultural Road, Vancouver, BC, V6T 1Z1, Canada\\
$^{2}$Institut de Ci\`encies del Cosmos, Universitat de Barcelona, Mart\'i Franqu\`es 1, E-08028 Barcelona, Spain\\
$^{3}$Astrophysics Research Institute, Liverpool John Moores University, 146 Brownlow Hill, Liverpool, L3 5RF, UK\\
$^{4}$Institute for Astronomy, University of Edinburgh, Royal Observatory, Blackford Hill, Edinburgh, EH9 3HJ, UK
}

\pubyear{2019}

\begin{document}
\label{firstpage}
\pagerange{\pageref{firstpage}--\pageref{lastpage}}
\maketitle

\begin{abstract}
The location of a galaxy cluster's centroid is typically derived from observations of the galactic and/or gas component of the cluster, but these typically deviate from the true centre.  This can produce bias when observations are combined to study average cluster properties. Using data from the BAHAMAS cosmological hydrodynamic simulations we study this bias in both two and three dimensions for 2000 clusters over the $10^{13} - 10^{15} ~\mathrm{M_{\odot}}$ mass range. We quantify and model the offset distributions between observationally-motivated centres and the `true' centre of the cluster, which is taken to be the most gravitationally bound particle measured in the simulation. We fit the cumulative distribution function of offsets with an exponential distribution and a Gamma distribution fit well with most of the centroid definitions. The galaxy-based centres can be seen to be divided into a mis-centred group and a well-centred group, with the well-centred group making up about $60\%$ of all the clusters. Gas-based centres are overall less scattered than galaxy-based centres. We also find a cluster-mass dependence of the offset distribution of gas-based centres, with generally larger offsets for smaller mass clusters. We then measure cluster density profiles centred at each choice of the centres and fit them with empirical models. Stacked, mis-centred density profiles fit to the Navarro-Frenk-White dark-matter profile and Komatsu-Seljak gas profile show that recovered shape and size parameters can significantly deviate from the true values. For the galaxy-based centres, this can lead to cluster masses being underestimated by up to $10\%$.
\end{abstract}




\section{Introduction}
\label{sec:Intro}

Galaxy clusters are the largest and most massive gravitationally-bound systems that we observe today, and as such, form a significant source of information to probe structure formation \citep{2019SSRv..215...24S,Walker2019} and constrain the cosmological parameters \citep{2010MNRAS.406.1759M,2011ASL.....4..204B}. Galaxy clusters are observed across the entire electromagnetic spectrum, and for that reason they also serve as a unique probe of the complex physical processes underlying galaxy formation and evolution. The future landscape of cluster science is very exciting, thanks to the development of new facilities, more sensitive equipment, and better resolution, allowing large areas of the sky to be surveyed across all wavelengths \citep{2019BAAS...51c.279M}. A key aspect of galaxy cluster science is the construction of cluster catalogues, which involves cluster identification and the measurement of basic features such as size, profile, redshift, morphology and mass. A crucial element of these measurements that is not often discussed is the definition of cluster centres.

Deconstruction of the cluster weak-lensing signal is a well developed and often used technique that allows us to study these clusters, most notably to recover their mass. However, in the process of this measurement, several assumptions about the shape and structure of the cluster have to be made. One of these is the choice of selecting a centre for the cluster. Traditionally, the definition for the `true' centre of a galaxy cluster is taken to be the deepest point of the gravitational potential well, i.e. where a test particle is most bounded to the system. However, calculating this requires knowing the mass distribution for the cluster in at least 2D (and ideally 3D including redshift) in the first place, which is difficult to do using the lensing signal alone. The work-around that has been employed for the past few decades is to use a host of different proxies for the centre that can be directly observed or measured. These proxies are based on our knowledge of the different physical components found in the galaxy cluster itself, such as hot X-ray emitting gas or luminous galaxies, and are theoretically well-motivated to trace the true centre. However, as they are fundamentally only approximations to the true cluster centre, our choice for the proxy centre invariably introduces a bias to our measurement of the cluster mass and shape. As the next generation of observing instruments such as the Large Synoptic Survey Telescope (LSST) reach precision capabilities where we can measure the mass to sub-percent levels, it becomes increasingly important to accurately quantify these centroid dependent biases and develop a systematic approach to correct for them.

Cluster density profiles are typically measured by stacking profiles from multiple sources, producing a convolution of the true mean profile with the probability distribution function (PDF) of the offset. A model is generally assumed for the PDF of the centroid offsets, and these model parameters are included when fitting \citep{cibirka2017codex}. In contrast, with cosmological simulations we know the position of the real centre of each galaxy cluster, thus we can constrain the offset PDF independently of the density measurements. In this study we use hydrodynamical simulations of galaxy clusters from the BAHAMAS project (BAryons and HAlos of MAssive Systems, \citet{McCarthy2017}) to measure and model the degree of mis-centring in galaxy clusters, and then analyse the effects this has on inferred cluster properties. In using the latest hydrodynamical simulations, we capture the significant effects that baryons have on the structure and formation history of the clusters, as opposed to studying dark matter only gravitational N-body simulations. Furthermore, as ideal environments where the full 3D distribution of the individual cluster components can be known, the hydrodynamical simulations allow us to exactly compute the various centroids, based on different tracers, for each cluster and compare them with the `true' centres. By studying a large number of clusters ($\sim 2000$) over a diverse mass range ($10^{13} - 10^{15} ~\mathrm{M_{\odot}}$), we can capture trends in centroid bias over different cluster shapes and sizes. 

These centroid-dependent biases are further explored in the study of the cluster mass density profiles, where empirically established models such as the Navarro-Frenk-White (NFW) dark matter profile \citep{NFW1997} and Komatsu-Seljak (KS) gas profile \citep{KS2001} can be fitted to data. From a theoretical point of view the profile fits inform us of how well current models capture the underlying mass distribution. More importantly, however, from an observational point of view, by performing the stacking of mis-centred profiles from multiple clusters we are able to study how the fitting parameters and related physical properties are affected. This is especially important, since identifying and using consistent centre proxies is essential for accurate stacking of cluster profiles to get high signal-to-noise when analysing real data.

Previous analyses of N-body simulations, as well as actual lensing data, showed that different centre choices can introduce biases of up to a few percent e.g. \citet{Schrabback2016}, \citet{Ford2015}, \citet{George2012}; see also the recent DES collaboration study \citet{Zhang2019}. \citet{cawthon2018dark} used the RedMaPPer cluster catalogue to characterise mis-centring, and they adopted the X-ray peak position as their reference cluster centre.  We extend their analysis with a larger (simulated) cluster catalogue, more centroid choices, and more offset distribution models.  

The article is organised as follows: in section~\ref{sec:Data} we first outline the data used from the BAHAMAS simulations along with the various centroid definitions and their calculations. In section~\ref{sec:Characterisation} we then quantify and model the degree of mis-centring. Section~\ref{sec:Profiles} follows with analysis of both well-centred and mis-centred density profiles. Key results are summarised in section~\ref{sec:Conclusions}.

\section{Data and Definitions}
\label{sec:Data}

\subsection{Simulation data}

\begin{figure*}
	\includegraphics[width=\textwidth] {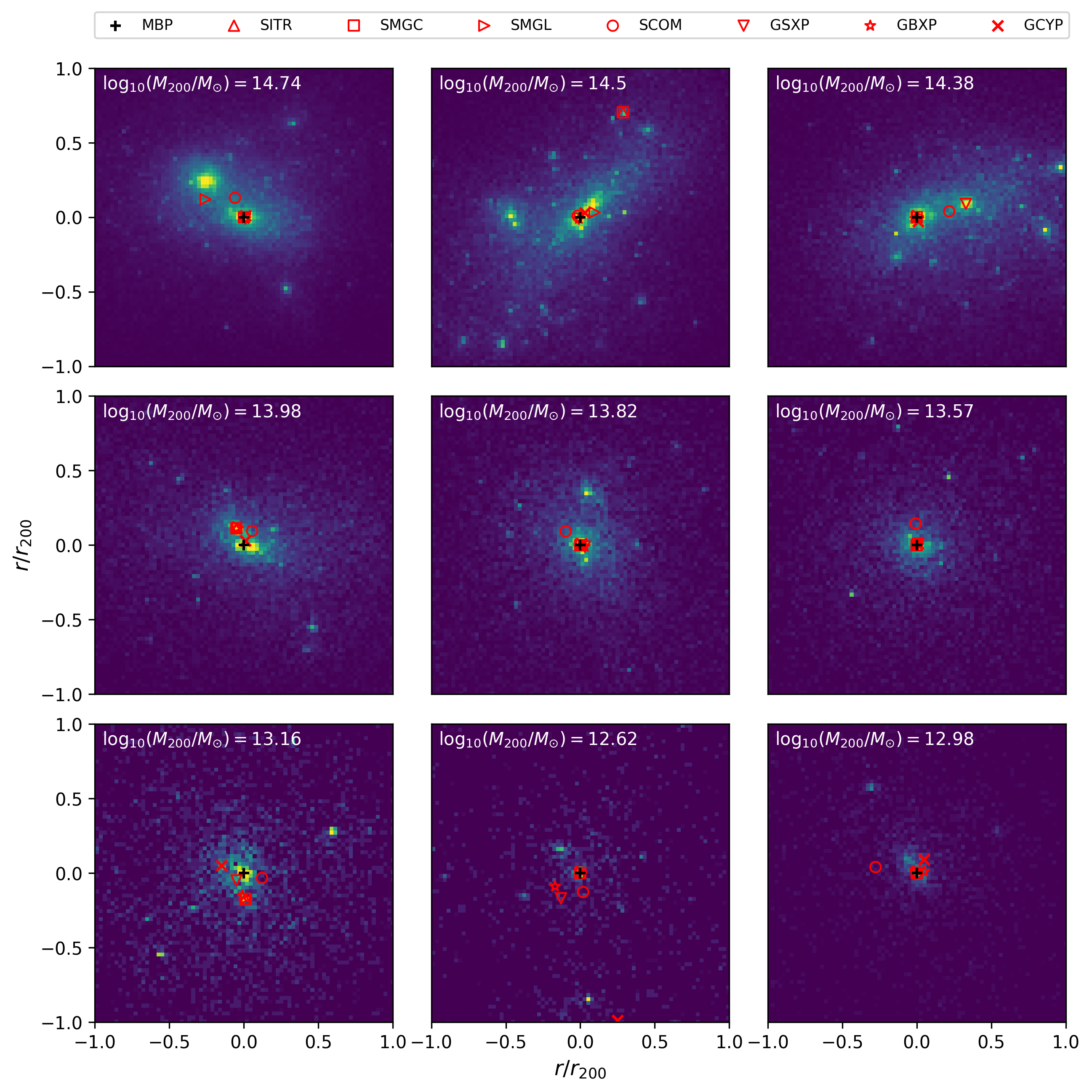}
    \caption{Images of 9 randomly selected clusters from the simulation colour-coded by stellar mass.  The projected 3D centroids defined in Table~\ref{table:centroid_def} are indicated by the symbols in the legend.}
    \label{fig:centroid_plot}
\end{figure*}

We employ data from the BAHAMAS (BAryons and Haloes of MAssive Systems, \citealt{McCarthy2017,McCarthy2018}) simulations.  BAHAMAS is a suite of cosmological, hydrodynamic simulations run using a modified version of the TreePM SPH code \texttt{GADGET3}.  The simulations primarily consist of $400~\mathrm{cMpc}/h$ periodic boxes containing $2 \times 1024^3$ particles (dark matter and baryonic, in equal numbers).  Here we use the run that adopts a WMAP 9-year best-fit cosmology with massless neutrinos \citep{Hinshaw2009}.

BAHAMAS includes subgrid treatments of important physical processes that cannot be directly resolved in the simulations, including metal-dependent radiative cooling, star formation, stellar evolution and mass-loss, black hole formation and growth, and stellar and active galactic nuclei (AGN) feedback. The subgrid models were developed as part of the OWLS project \citep{Schaye2010}. The parameters governing the efficiencies of AGN and stellar feedback were adjusted so that the simulations approximately reproduce the observed galaxy stellar mass function for $\mathrm{M_*} \geq 10^{10}~\mathrm{M_{\odot}}$ and the hot gas fraction--halo mass relation of groups and clusters, as determined from high-resolution X-ray observations of local systems.  As shown in \citet{McCarthy2017}, the simulations match the galaxy--halo--tSZ--X-ray scaling relations of galaxies and groups and clusters.

For the present study, friends-of-friends (FOF) haloes are selected in logarithmic bins of $M_{200,{\rm crit}}$ from a reference dark matter-only simulation\footnote{We selected haloes from a reference dark matter-only simulation, so as to facilitate future comparisons with runs that vary feedback and cosmology.}, which are then matched to the fiducial BAHAMAS hydro simulation.  We adopt a bin width of $0.25$ dex spanning the range $10^{13} - 10^{15} ~\mathrm{M_{\odot}}$ and randomly select 300 unique haloes per mass bin (or all of the haloes, if there are fewer than 300 in a bin), resulting in a sample of approximately $2000$ haloes.

All particles (representing gas, dark matter and stellar content) in a sphere of radius $2 r_{200}$ centred on the most bound particle (MBP) are selected for analysis.  In addition, a galaxy catalogue is produced for all simulated galaxies with $M_{gal} > 10^{10} ~\mathrm{M_{\odot}}$ within this radius.  Simulated galaxies are defined as the stellar component of self-gravitating substructures identified with the \texttt{SUBFIND} algorithm.

\subsection{Halo centre definitions}

For this study, the halo centre is defined to be the position of the gravitationally most bound particle (MBP), i.e. the halo particle with the minimum gravitational potential.  An alternative `true centre' is the point where the total mass density is maximum. As it happens, this coincides with the MBP for most clusters, and since neither this nor the MBP are directly observable, we arbitrarily choose the former as the {\it true} centre, both in 2D and 3D. We then study the effect that choosing each of seven alternative cluster centre definitions has on the inferred cluster mass and density profile, both in 2D and 3D. These proxy centres are based on different cluster components, which in turn are close to physical cluster observables. The names, definitions and abbreviations for these centroids are given in Table \ref{table:centroid_def}. These proxy centres can be coarsely classified into gas-based or stellar-based tracers.

\begin{table*}
\caption{The names, abbreviations and component tracers of the centroids analysed in this study. The Most Bound Particle (MBP) is taken to be the true centre of the galaxy cluster, and the seven other centroids are evaluated with respect to it.}
\begin{tabular}{lll}\hline\hline
Abbreviation & Full Name                                      & Tracer           \\
\hline
MBP          & Most Bound Particle                            & -                \\
{SITR}        & Stars Centre from Iteration Method                    & Galaxies       \\
{SMGC}        & Stars Most-Massive Galaxy Centre                     & Galaxies         \\
{SMGL}	     & Stars Median Galaxy Location					  & Galaxies         \\
{SCOM}         & Stars Centre Of Mass                        & Galaxies         \\
GSXP       & Gas Soft X-ray luminosity Peak          & Soft X-ray       \\
GBXP       &  Gas Bolometric X-ray luminosity Peak   & Bolometric X-ray \\
GCYP     &  Gas Compton $y$ parameter Peak & Compton $y$       \\
\hline\hline
\end{tabular}
\label{table:centroid_def}
\end{table*}

The gas-based proxies, GSXP, GBXP and GCYP, are defined as the peak position of the soft X-ray, bolometric X-ray, and Compton $y$ parameter luminosity, respectively. For each gas particle in the simulations, we have an associated luminosity for each type of emission. To calculate the centroids we first pixelize all the gas particles onto a grid with a pixel/voxel size of 0.01$r_{200}$, then we smooth the luminosities with a Gaussian beam whose width is equal to the mean gas inter-particle distance. The centroids are defined as the centre of the pixel/voxel with the highest luminosity for each of the three observables. 

Stellar-based centroids are calculated from the BAHAMAS galaxy catalogue, which is based on stellar mass distributions. The Galactic Centre from Iteration Method (SITR) is based on a method described in \citet{10.1111/j.1365-2966.2011.19217.x}. For each cluster, we first calculate the centre of stellar mass, then remove the galaxy that is furthest from this centre. In each subsequent iteration, the new centre of stellar mass is found and the furthest galaxy is once again dropped, until there are only two galaxies left. SITR is then taken to be the centre of the more massive galaxy. The Most Massive Galaxy Centre (SMGC) is simply the position of the galaxy with the highest mass inside the cluster halo. The Galactic Centre of Mass (SCOM) is the galaxy-mass-weighted mean position of objects within the halo. The Median Galaxy Count Location (SMGL) \citep{andreon2015making} is another iterative scheme.  For each cluster, we start with a $(2r_{200})^3$ volume centred at the SMGC. We count galaxies along each coordinate direction from the origin and find the median position along each axis.  We move the centre to the new median position and clip the volume to $0.85$ times the previous volume and repeat the search. The iteration ends when the volume dimension reaches $0.4r_{200}$.  The SMGL is taken to be the position from the last iteration.

For our 2D analysis we generate data from the simulation using the distant-observer approximation, by projecting the positions of all cluster particles onto the plane of the sky.  The calculation of each centroid follows the same definition as in 3D.  The 2D centres calculated this way more closely resemble real data, while the 3D centres allow us to utilise the full information available from the simulation.

Fig.~\ref{fig:centroid_plot} shows nine randomly selected clusters from the simulation, with the centroids listed in Table~\ref{table:centroid_def} highlighted. 

\section{Characterisation of mis-centring}
\label{sec:Characterisation}

\subsection{Modelling the Offset Distribution}
\label{subsec:model_offset}
\begin{figure}
\centering
	\includegraphics[width=0.4\textwidth] {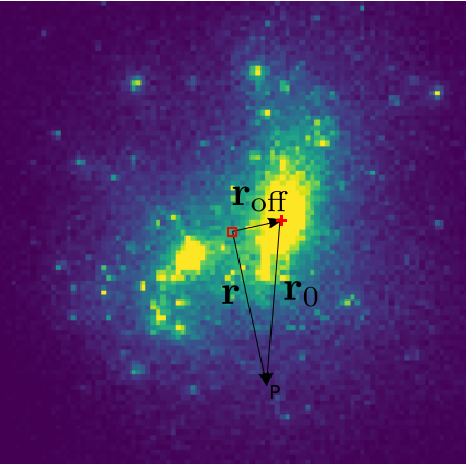}
    \caption{An example illustrating the relation between $\textbf{r}$, $\textbf{r}_{\mathrm{off}}$ and $\textbf{r}_0$. $P$ is the point where the density profile is measured. The red square is the centroid and red cross is the MBP.}
    \label{fig:cluster_offcentre}
\end{figure}

{
The stacked density profile of a group of mis-centred galaxy clusters is typically modelled as an astrophysical profile convolved with a Probability Distribution Function (PDF) of the offset:
\begin{equation}
    \rho_{\mathrm{off}}(r|\{q_i, s_j\}) = \int \rho(r_0|\{q_i\})\mathrm{P}(r_{\mathrm{off}}|\{s_j\})\mathrm{d}\mathbf{r}_{\mathrm{off}},
    \label{eq:off_profile}
\end{equation}
where the centroid offset vector $\bf{r}_{\rm off}$ is defined as:
\begin{equation}
\bf{r}_{\rm off}\equiv {\bf r}_{\rm MBP}-{\bf r}_{\rm cen},
\end{equation}
where ${\bf r}_{\rm cen}$ is one of the centroid positions defined in Table~\ref{table:centroid_def} (in 2D or 3D, we use bold type to represent vectors and unbold type to represent their magnitude). In the above, ${\bf r}$ is the position vector with respect to the measured centre, while ${\bf r}_0$ is the position vector relative to the MBP (the true centre):
\begin{equation}
\textbf{r}_0 \equiv \textbf{r} - \ \bf{r}_{\rm off},
\end{equation}
The relation between $\textbf{r}$, $\textbf{r}_{\mathrm{off}}$ and $\textbf{r}_0$ is illustrated in Fig.\ref{fig:cluster_offcentre}.

In Eq.~\eqref{eq:off_profile}, $\mathrm{P}(r_{\mathrm{off}}|\{s_j\})$ is the PDF model parameterized by $\{s_j\}$. So the average profile of mis-centred clusters is described by the density profile parameters $\{q_i\}$ and the offset PDF model parameters $\{s_j\}$. With real data, both $\{q_i\}$ and $\{s_j\}$ must be constrained simultaneously \citep{cibirka2017codex}. In this paper we constrain the $\{s_j\}$ separately, since we know the MBP positions from the simulation.}

\begin{figure*}
	\includegraphics[width=\textwidth] {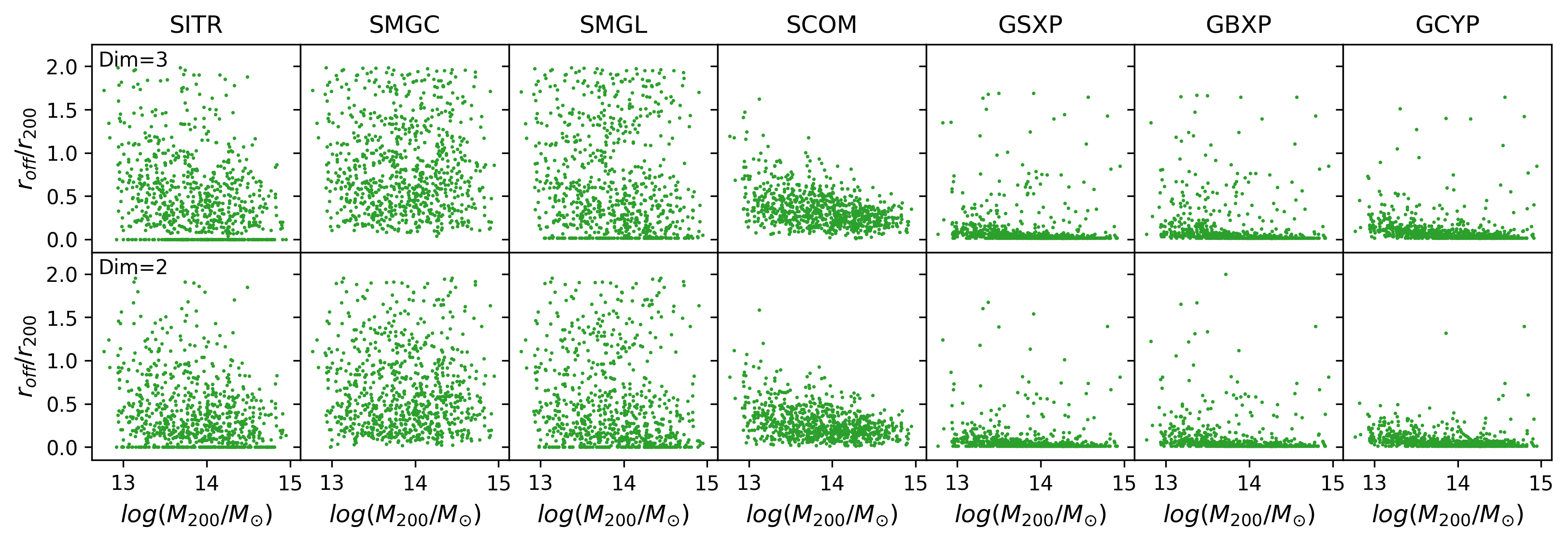}
    \caption{Centroid offsets with respect to the MBP as a function of cluster mass, $M_{200}$, for each centroid defined in Table~\ref{table:centroid_def}. The upper panels show results for the 3D centroids, the lower panels for the 2D centroids. For the first three galaxy-based centroids, a significant fraction of the clusters have centroids that coincide with the MBP, while the rest show significantly higher scatter than the gas-based centroids possess.}
    \label{fig:moff_scatter}
\end{figure*}

In this paper, the offsets are measured in units of $r_{200}$. Fig.~\ref{fig:moff_scatter} shows the distribution of offsets with respect to $M_{200}$. In the figure, we see that the offsets for some of the galaxy-based centroids split into two groups. For these centres, there is one population that effectively tracks the MBP, while the other centres are quite scattered. In order to capture this behaviour, we form a bimodal offset model. {\citet{cawthon2018dark} find that a combination of an Exponential distribution and a Gamma distribution (denoted E$\Gamma$ below) is the best model to describe their distribution of centroid offsets. We adopt this model in our analysis, but we consider other parametrizations in Appendix~\ref{sec:app}.}

The specific form of the E$\Gamma$ model is:
\begin{equation}
\begin{aligned} \mathrm{P_M}\,(r | f, \sigma, \tau) = f \times\frac{1}{\sigma}\exp\left(-\frac{r}{\sigma}\right)+(1-f) \times \frac{r}{\tau^2}\exp\left(-\frac{r}{\tau}\right) ~,
\end{aligned}
\label{eq:pdf_model}
\end{equation}
where $\sigma$ and $\tau$ are parameters that describe the width of each population and $f \in [0, 1]$ is the fraction of clusters that belong to the centred population.

\begin{figure*}
\subfloat{
	\includegraphics[width=\textwidth] {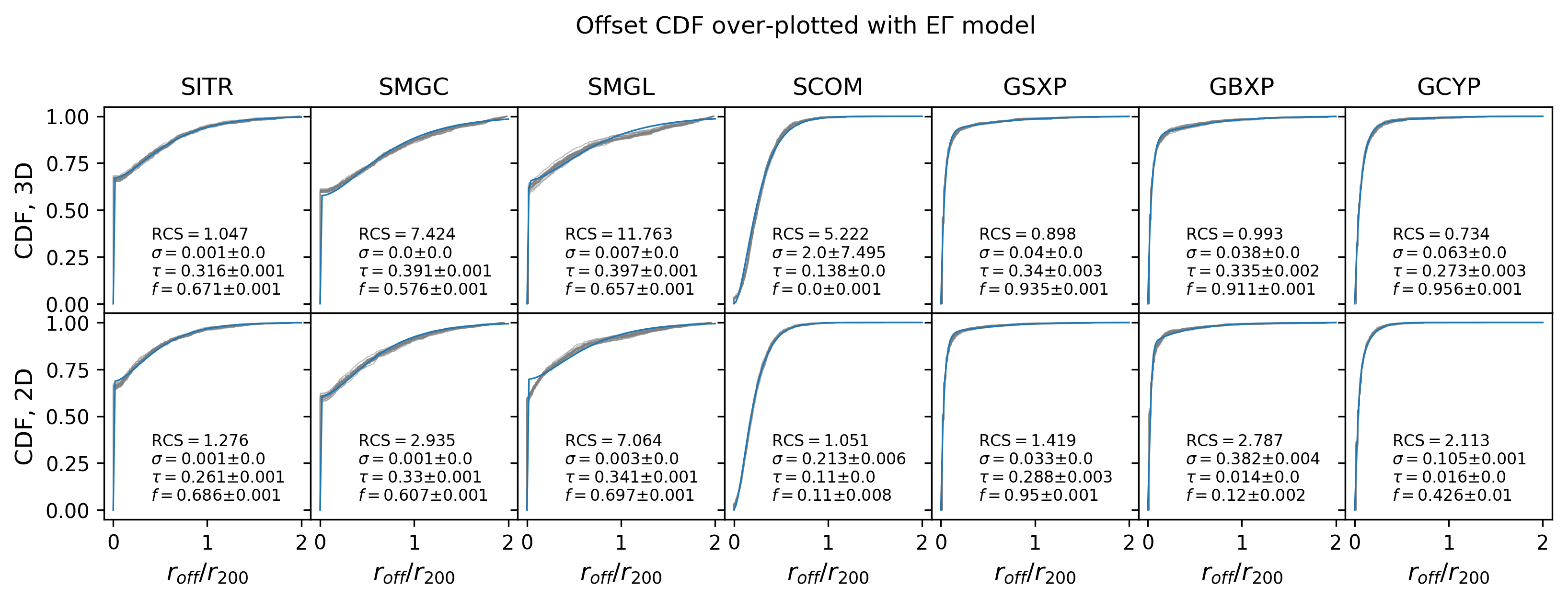}}
    \caption{The measured (grey) offset CDFs for all the seven centroid definitions, along with their corresponding best-fit E$\Gamma$ model distributions. The best-fit model parameters and reduced $\chi^2$ (RCS) values are indicated in each panel. Upper panels: 3D offsets; lower panels: 2D offsets.}
    \label{fig:cdf_bestfit}

\end{figure*}

In order to avoid binning artefacts, we fit the model parameters ${f, \sigma ~\mathrm{and}~ \tau}$ to the cumulative distribution function (CDF) given by,
\begin{equation}
\begin{aligned}
\mathrm{C_M}\,(r_{\rm off}|f, \sigma, \tau) = \int_0^{r_{\rm off}}\mathrm{P_M}\,(r|f, \sigma, \tau) ~\mathrm{d}r\ .
\end{aligned}
\label{eq:cdf_model}
\end{equation}
For each centroid definition, we estimate an empirical $\mathrm{C_M}(r_{\rm off})$ from the data using all 2000 clusters.  We use Bootstrap methods to estimate the covariance of each CDF by re-sampling the clusters 200 times using random draws with replacement. From this ensemble, we calculate the mean ${\rm C_M}$ and its covariance, then we fit the model in Eqs. \eqref{eq:pdf_model} and \eqref{eq:cdf_model}. When fitting, we constrain $\sigma, \tau \in [0,2]$ and $f \in [0,1]$. Fig. \ref{fig:cdf_bestfit} shows our sampled CDF estimates along with our model.  

The best-fit E$\Gamma$ model is compared to the measured CDF in Figs.~\ref{fig:cdf_bestfit}. The measured CDF clearly shows the bimodal behaviour of the SITR, SMGC and SMGL centroids.  Note that $\sigma$ in these cases is much smaller than $\tau$ or $r_{200}$, which means that the first group, defined by the Exponential distribution, is the well-centred group.  This group comprises roughly 60\% of the clusters. The SITR centroid has a slightly higher well-centred fraction, which means that the iteration method rules out some massive galaxies that are not physically part of the cluster. For the well-centred group, the most massive galaxy is at the minimum of the gravitational potential of its host cluster.

For the stars centre of mass and gas-based centroids (SCOM, GSXP, GBXP and GCYP), it is inappropriate to interpret the distribution in terms of a well-centred and mis-centred group because the distributions are clearly not bimodal. In these cases, the Exponential and Gamma distributions are simply parameterizing one underlying distribution with additional degrees of freedom.  In most cases, however, one model dominates over the other, as measured by the fraction, $f$.

In summary, our results are mixed. When considering all clusters, the gas-based centroids have a smaller scatter than the galaxy-based ones, but the latter centroids significantly out-perform the former in roughly 60\% of the clusters. Unfortunately, {it is difficult to distinguish whether a given cluster is well-centred or mis-centred from the available data. We are currently investigating whether new tools, such as machine learning algorithms, might be of use in addressing this question. 
 Amongst the galaxy-based centroids, the iterative SITR definition has the highest well-centred fraction and the lowest PDF width, while amongst the gas-based centroids, the X-ray based ones produce slightly smaller overall scatter than the Compton $y$.

We repeated the above analysis with several different PDF parametrizations (see Appendix~\ref{sec:app}) and concluded that the E$\Gamma$ model is adequate to parameterize the PDF of the centroid offset distribution.  This conclusion agrees with \citet{cawthon2018dark}.} 

\subsection{Mass Dependence of the Offsets}

\begin{figure}
\includegraphics[width=\columnwidth]{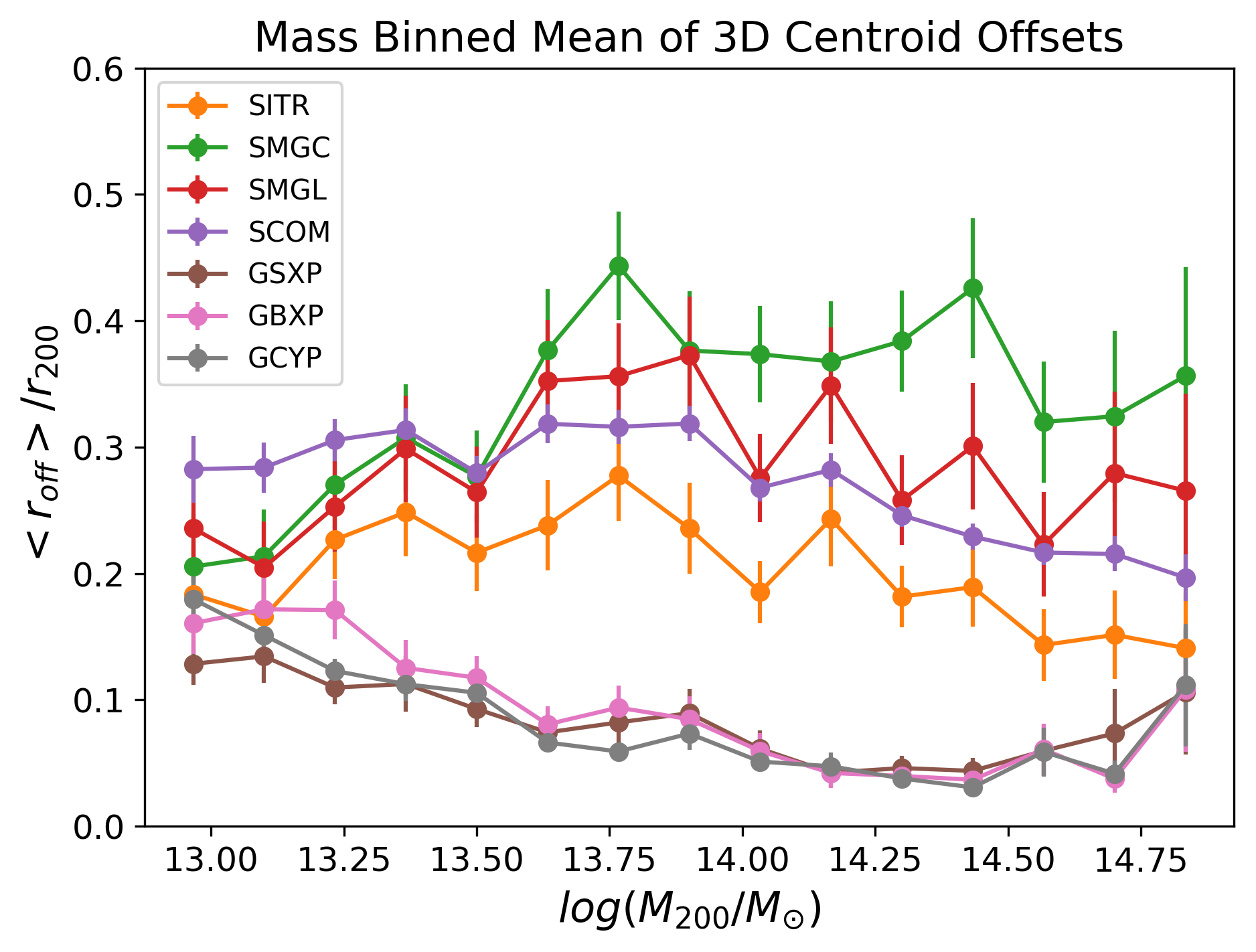}
    \caption{The mean 3D centroid offset as a function of cluster mass for each centroid definition. The clusters are binned by their value of $M_{200}$ between from $10^{13}$ and $10^{15} ~\mathrm{M_{\odot}}$. The plotted uncertainty is the standard deviation of the centroid offset in each mass bin. The 2D results are qualitatively similar.}
    \label{fig:3d_mean}
\end{figure}

We next examine the mass dependence of the various centroid offsets.  To do so, we segregate the clusters into 15 mass bins of width 0.13 dex, from $M_{200} = 10^{13}$ to $10^{15} ~\mathrm{M_{\odot}}$, and then compute the mean offset for each centroid in each mass bin. The results for our 3D analysis are shown in Fig.~\ref{fig:3d_mean}; the 2D results are qualitatively similar.


The results show that the galaxy-based centroids SITR, SMGC and SMGL generally have larger mean offsets than the gas-based centroids. The SCOM offsets hover around $0.3\,r_{200}$ until a cluster mass of about $10^{14} ~\mathrm{M_{\odot}}$, above which they steadily decreases with mass. This may occur because the more massive clusters contain a larger number of galaxies with $M_{\mathrm{gal}} > 10^{10} ~\mathrm{M_{\odot}}$, which may reduce the shot noise due to outlying massive galaxies that are far from the MBP. The gas-based centroid offsets exhibit a more systematic decrease with cluster mass until $\sim 10^{14.4} ~\mathrm{M_{\odot}}$, above which they start to increase.  This latter effect may be due to high-mass clusters being more likely to have undergone a major merger in the past, and therefore to have more gas substructure their halos \citep{mihos2003interactions}.

\section{Mass Density Profiles}
\label{sec:Profiles}

Mass density profiles provide valuable information about the size and structure of a galaxy cluster.  However, it is usually necessary to average multiple cluster profiles together to gain high signal to noise.  In doing so, the chosen cluster centre becomes important, as mis-centring produces biased profiles which lead to biased inferences of their physical properties \citep{George2012}. Our aim here is to investigate and quantify this bias as a function of centroid definition.

There are two ways of stacking profiles from cluster observations. One is called the `observer point of view', in which cluster profiles with a similar mass are first stacked and averaged, then the average profile is fit to a model. The other is called the `theorist point of view', in which each cluster profile is individually fit to a model, without stacking, after which the fit parameters are averaged. Since the latter approach requires high signal to noise data on individual clusters, we do not explore it further in this paper, choosing instead to characterise the `observer point of view'.

\subsection{Methods and Definitions}

In this section, we generate average mass density profiles (in 2D and 3D) as a function of cluster mass for each of the centroid definitions presented in Table~\ref{table:centroid_def}.  The clusters are grouped into 10 mass bins of 0.2 dex each, in the range $10^{13} - 10^{15} ~\mathrm{M_{\odot}}$, and the profiles are binned in units of $r_{200}$, defined with respect to the MBP.  For diagnostic purposes, we generate density profiles including {\it all} particles in a cluster, and profiles that includes only gas particles (see below for details). In all, we produce 32 density profiles for each cluster (8 centroid definitions $\times$ 2D or 3D $\times$ gas or all particles), before forming the average profiles.

To form a density profile, we group a cluster's particles into 20 equal log-spaced radial bins in the range $0.1\,r_{200} < r < r_{200}$ according to the particle's distance to the chosen centroid. The bin size and spacing provides a reasonable balance between resolving profile structure and minimising shot noise within each bin. The 3D profile is measured by summing the masses of particles within each bin and dividing by the bin volume,
\begin{equation}
\rho(\bar{r_i}) = \frac{\sum_{a}M_a}{\frac{4\pi}{3}\left(r_{i+1}^3-r_{i}^3\right)}\, .
\label{eq:3d_profile}
\end{equation}
Here, $r_{i}$ is the inner edge of the $i$-th bin, $\bar{r_i}\equiv(r_{i}+r_{i+1})/2$ is the bin centre, and $M_a$ is the mass of $a$-th particle within the $i$-th radial bin.  Note that the sum on $a$ will either include {\it all} particles in the bin, or only the gas particles.

The 2D profile is defined similarly, except that $r$ is a projected distance, and the denominator is the bin area,
\begin{equation}
\Sigma(\bar{r_i}) = \frac{\sum_{a}M_a}{\pi\left(r_{i+1}^2-r_{i}^2\right)}
\label{eq:2d_profile}
\end{equation} 
Once these profiles are in hand for each cluster and configuration, we average the $\sim 200$ cluster profiles in each mass bin, and calculate the bin-bin covariance matrix. We then fit the stacked profiles with empirical models from literature, as detailed below.

We model the ``all-particle'' density profiles with the Navarro-Frenk-White (NFW) model, based on dark-matter-only N-body simulations \citep{NFW1997}. This form is commonly used to describe cluster density profiles (see, for example, \citet{lokas2001properties}),
\begin{equation}
\rho_{\mathrm{NFW}}(r) = \frac{\rho_0}{\frac{r}{R_s}\left(1+\frac{r}{R_s}\right)^2}
\label{eq:nfw_model}
\end{equation}
where $R_{\mathrm{s}}\equiv c\,R_{200}$ is the scale radius and $c$ is a dimensionless concentration parameter.

For the gas we use the Komatsu-Seljak (KS) model which derives from early hydrodynamical simulations \citep{KS2001},
\begin{equation}
\rho_{\mathrm{KS}}(r) = \rho_0 \left(\frac{\ln\left(1+\frac{r}{R_s}\right)}{\frac{r}{R_s}}\right)^{1/\Gamma - 1}
\label{eq:ks_model}
\end{equation}
where $R_s$ is as defined in the NFW model, $\rho_0$ is the central density parameter, and $\Gamma$ is the polytropic index of the gas, which defines the gas' equation of state, $P_{\mathrm{gas}}\propto\rho_{\mathrm{gas}}^{\Gamma}$, where $P_{\mathrm{gas}}$ is the gas pressure and $\rho_{\mathrm{gas}}$ is the gas density.

For 2D profile fitting, the above models are projected according to
\begin{equation}
\Sigma (r) = 2\int_0^{\sqrt{r_{200}^2-r^2}}\rho(\sqrt{r^2+z^2})\, \mathrm{d}z,
\end{equation}
where the upper limit of integration is not $\infty$ because we only measure profiles for particles within $r_{200}$ (in 2D and 3D).

For the NFW fits, we take $c$ and $R_s$ as the free parameters. For the KS fits, we first fit the corresponding dark-matter-only profile to the NFW model to get $R_s$, then we fit the gas profile to the KS model with $\rho_0$ and $\Gamma$ as free parameters. The reduced $\chi^2$ values for each fit are also tracked as a measure of goodness-of-fit. Once the best-fit parameters are determined, we calculate the mass bias for each mass bin and centroid definition (Table~\ref{table:centroid_def}), defined as
\begin{equation}
\mathrm{Mass \ bias} \equiv \left( \frac{M_{200}^{\rm fit}}{M_{200}^{\rm MBP}} - 1\right)\times 100
\end{equation}
where $M_{200}^{\rm fit}$ is
\begin{equation}
M_{200}^{\rm fit} \equiv \int_0^{r_{200}}\rho(r|\{p_\mathrm{best~fit}\})\,4\pi r^2{\rm d}r.
\end{equation}
Here $\{p_{\rm best~fit}\}$ is the set of best-fit parameters: for the all-particle case it is $\{\rho_0,~c,~R_s\}$, while for the gas-only case it is $\{R_s,~\rho_0,~\Gamma\}$. Note that for the 2D analysis, $M_{200,\mathrm{fit}}$ is obtained from the 3D profile, $\rho(r)$, but with profile parameters that are fit to the 2D data, $\Sigma(r)$.  The ``true'' value, $M_{200}^{\rm MBP}$, is measured by directly summing the masses of all particles within $r_{200}$ of the MBP. In order to estimate uncertainties in the fit parameters, we bootstrap re-sample clusters in each mass bin 20 times, with replacement. Each time we stack and fit for the model parameters, evaluate $\chi^2$, and measure the mass bias. The uncertainties are given by the standard deviation of the re-sampled estimates.

\subsection{Centred Fit Analysis}
\label{sec:centred}

\begin{figure*}
\includegraphics[width=0.8\textwidth]{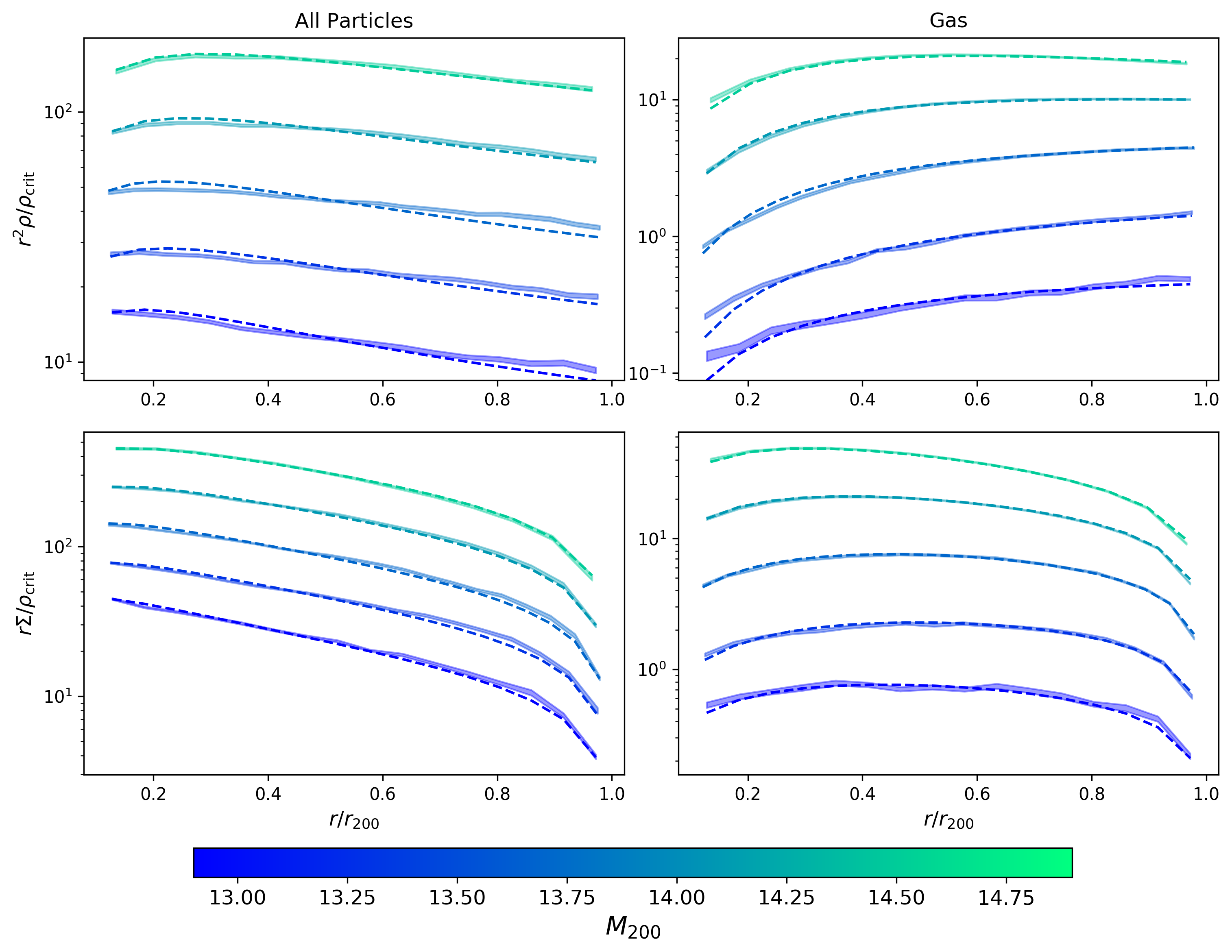}
    \caption{Stacked cluster density profiles, centred on the MBP (the true centre). The left panels show the all-particle profiles along with the best-fit NFW profiles (dashed); the right panels show the gas density profiles, along with the best-fit KS profiles (dashed). Upper panels show the 3D profiles while the lower panels show the 2D surface density profiles. Different colours correspond to different mass bins (only a selection of mass bins are shown.)  The small model errors due to stacking produce a small mass bias, as shown in Fig.~\ref{fig:3d_profile_param}.}
    \label{fig:MBP_profiles_3d}
\end{figure*}

To establish a baseline, we first conduct a profile analysis relative to the ``true'' centre, the MBP. The 3D and 2D profiles and fits are shown in Fig.~\ref{fig:MBP_profiles_3d}.  The stacked cluster profiles are not especially well described by the models, which is reflected in reduced $\chi^2$ value of $\sim$2-3, as shown in Fig.~\ref{fig:3d_profile_param}. This is due to the fact that the average density profiles are formed from clusters of different masses (within the mass bin). The discrepancy is resolved if we adopt the theorist perspective: individually fitting each cluster profile, then averaging the fit parameters within each mass bin. However, since we are trying to characterise the biases that result from stacking profiles, the mis-centred results presented below include the small bias due to stacking over a finite mass range.

In both the all-particle case (fit by the NFW model) and the gas case (fit by the KS model), the recovered mass is underestimated by a few percent. As shown in the lower right panels of Fig.~\ref{fig:3d_profile_param} (blue curves), the bias is largest for the lighter clusters, and approaches zero for the most massive clusters.

\subsection{Mis-centred Fit Analysis}

\begin{figure*}
\subfloat{
\includegraphics[width=0.5\textwidth]{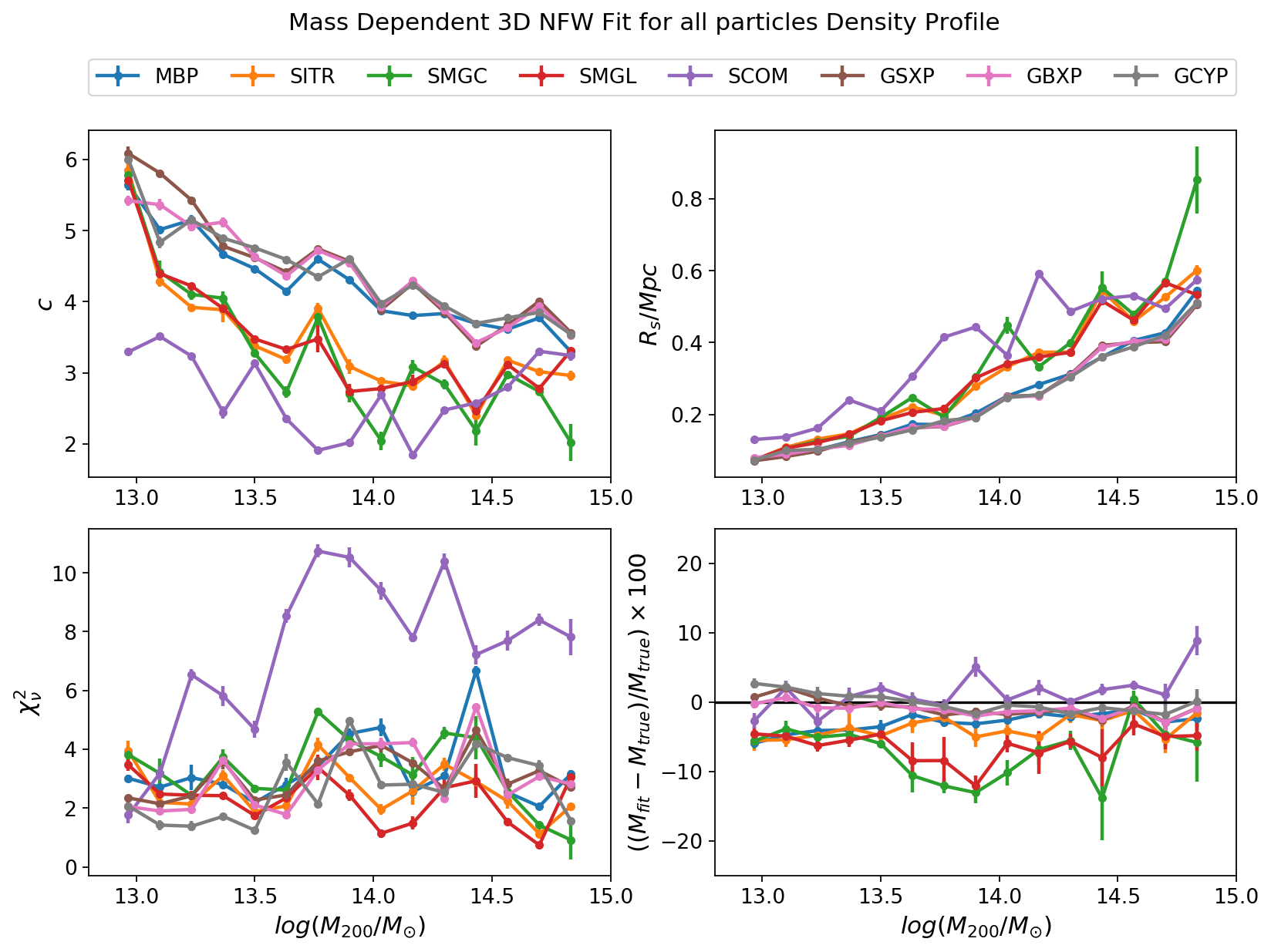}}
\subfloat{
\includegraphics[width=0.5\textwidth]{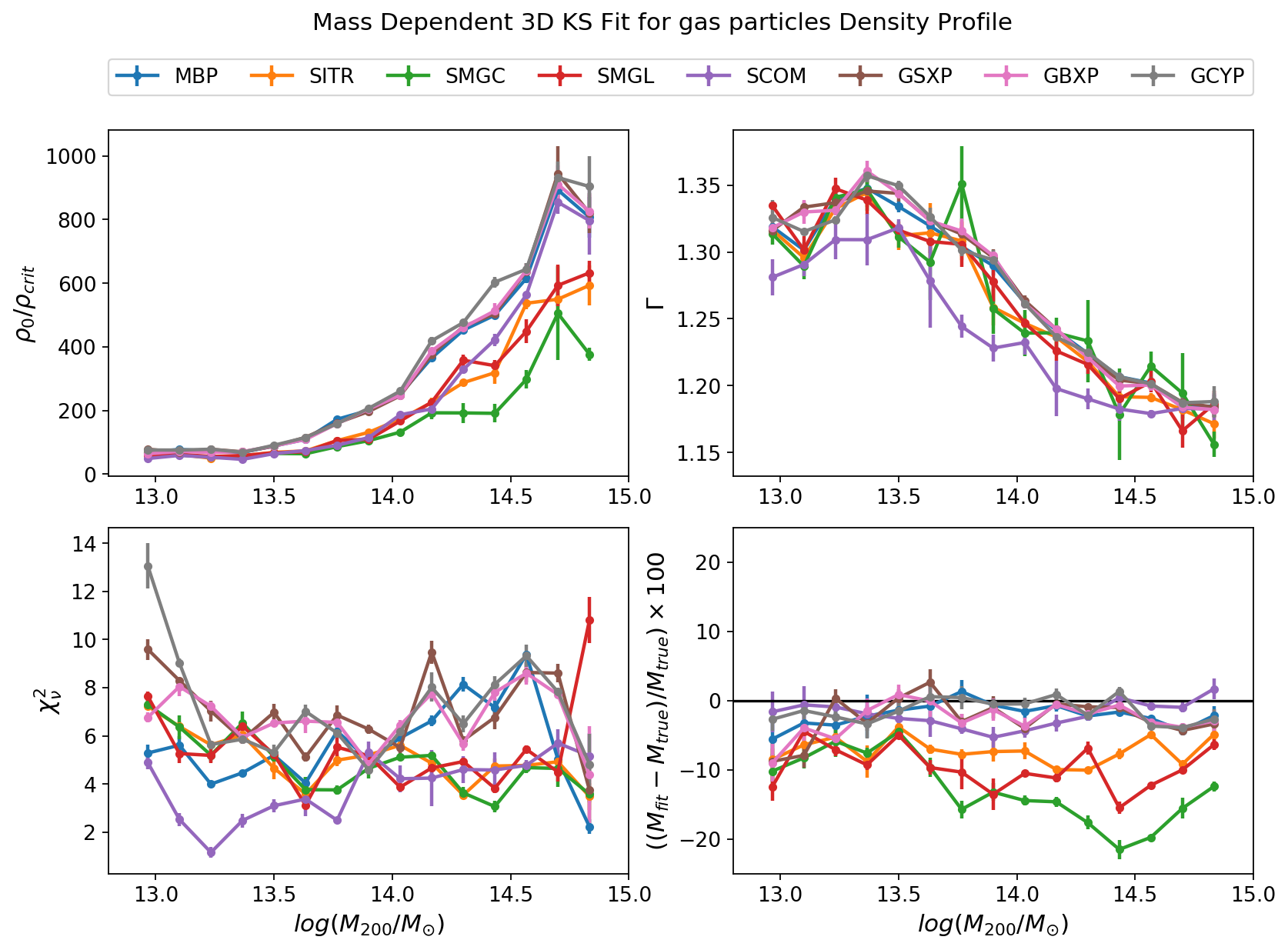}}
    \caption{Left panels: NFW model parameters fit to the 3-D all-particle density profiles. Right panels: KS model parameters fit to the 3-D gas density profiles. For each case, the upper panels show the best-fit model parameters while the lower panels show the reduced $\chi^2$ of each fit and the inferred mass bias.}
    \label{fig:3d_profile_param}
\end{figure*}

\begin{figure*}
\subfloat{
\includegraphics[width=0.5\textwidth]{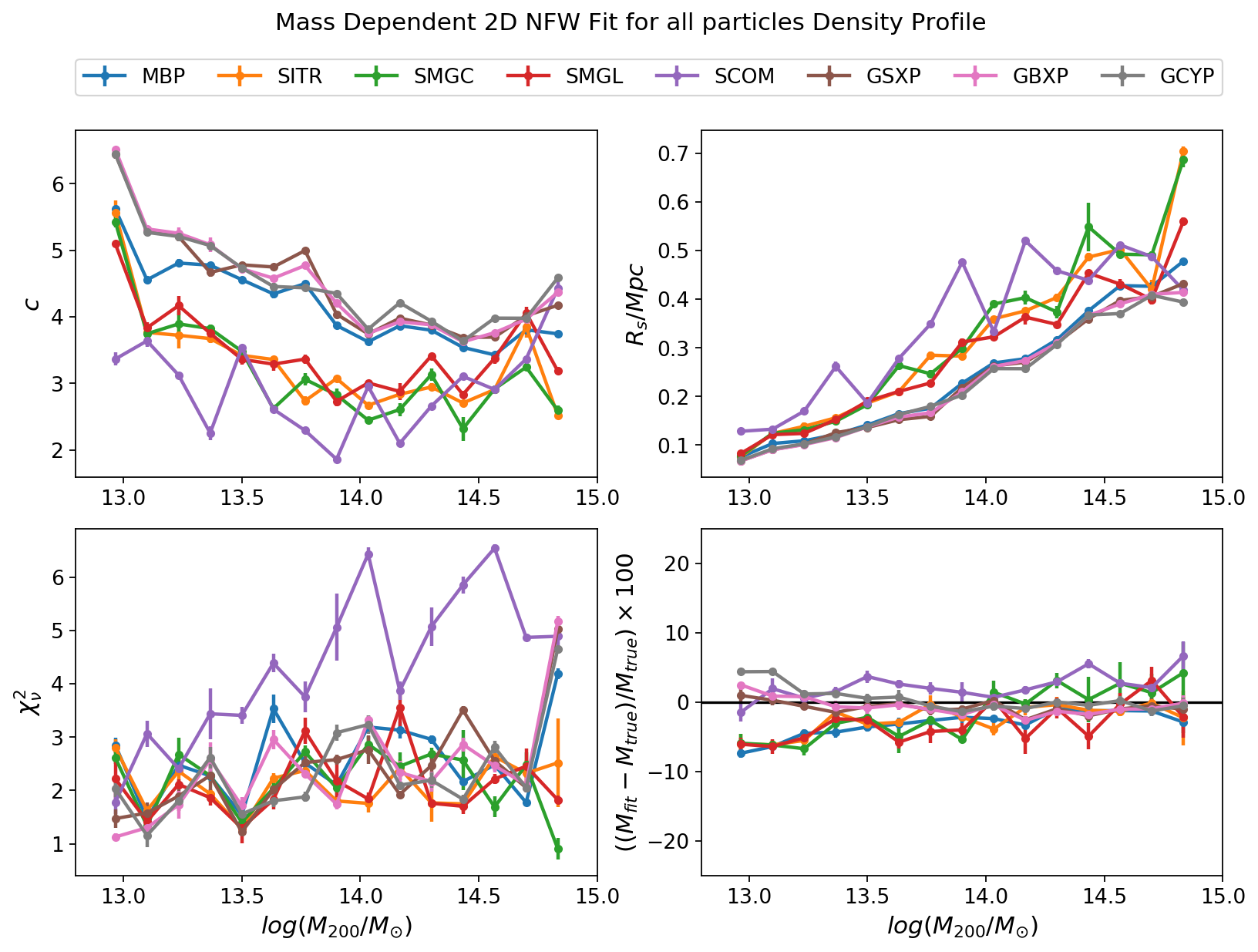}}
\subfloat{
\includegraphics[width=0.5\textwidth]{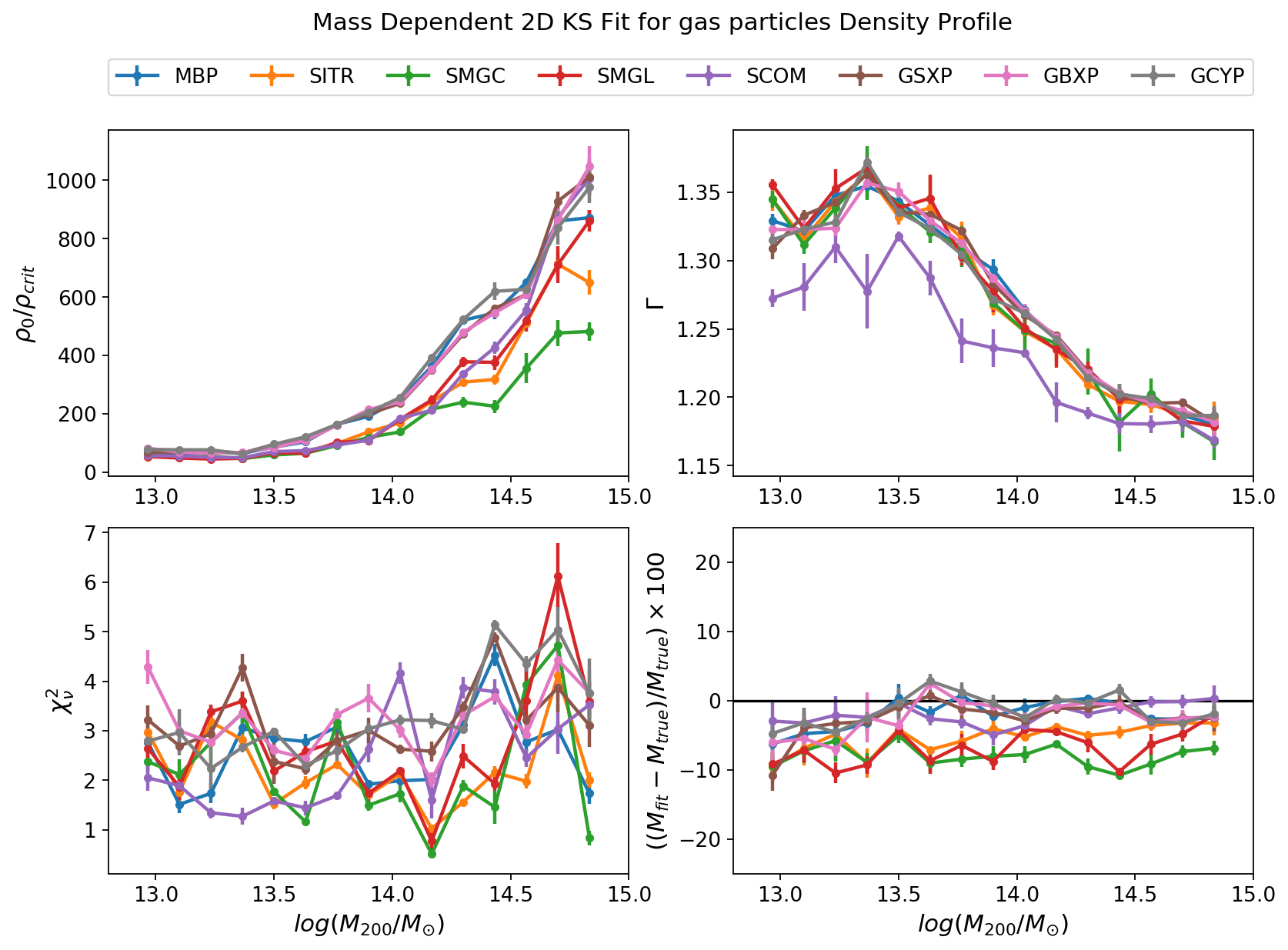}}
    \caption{Left panels: NFW model parameters fit to the 2-D all-particle density profiles. Right panels: KS model parameters fit to the 2-D gas density profiles. For each case, the upper panels show the best-fit model parameters while the lower panels show the reduced $\chi^2$ of each fit and the inferred mass bias.}
    \label{fig:2d_profile_param}
\end{figure*}

The best-fit profile parameters for each centroid definition in Table~\ref{table:centroid_def} are shown in Figs.~\ref{fig:3d_profile_param} and \ref{fig:2d_profile_param}.  The results are plotted as a function of the true cluster mass, $M_{200}$, measured with respect to to the MBP. For each case (NFW and KS), the upper panels show the best-fit model parameters while the bottom panels shows the reduced $\chi^2$ for each fit along with the inferred mass bias.   Uncertainties are estimated using bootstrap re-sampling.

For the NFW profiles, the 3 gas-based centroids generally yield small mass biases of $<\sim 5\%$, in accord with their small offsets. The SCOM-centred profiles fit badly but the mass error ends up being relatively small. It appears that shifts in the fits of $c$ and $R_s$ (relative to the MBP-based values) compensate in the product $R_{200}\equiv c\,R_{s}$, leaving only a small bias in $M_{200}$.  The SMGC and SMGL-centred profiles produce mass biases of up to 10-20\% in the 3D case but less than 10\% in 2D.  The SITR-centred profiles produce a mass bias of $\sim$6\% in 3D, which agrees with our earlier observation that SITR is a marginally better galaxy-based centroid, comparing to SMGC and SMGL.

For the KS profiles, the MBP, SCOM, and the three gas-based centroids generally provide unbiased cluster masses but the SMGC and SMGL centroids produce a relatively large mass bias of $\sim$10-15\% in 3D and $\sim$8\% in 2D. The SITR-centred profiles give a slightly smaller mass bias than the SMGC and SMGL centroids, especially for high-mass clusters. Overall, there are no compelling trends as a function of the true cluster mass.

As a final note, we explore the degree to which stacking itself, rather than mis-centring, is contributing to bias in the inferred parameters.  To do so, we measure and fit the density profiles for each cluster, then calculate the mean and standard deviation of the best-fit model parameters within a mass bin. The reduced $\chi^2$ is significantly closer to 1, in agreement with the MBP-based results in \S\ref{sec:centred}. However, the average best-fit model parameters and inferred mass biases are not significantly different from the results of fitting to the stacked profiles.  We conclude that the biases shown in Figs.~\ref{fig:3d_profile_param} and \ref{fig:2d_profile_param} are indeed due to centroid offsets.


\section{Conclusions}
\label{sec:Conclusions}

Using catalogues derived from the BAHAMAS hydrodynamical simulations, we investigate seven observationally motivated cluster centroid definitions (Table~\ref{table:centroid_def}). These include: the star centre from iteration method (SITR), the star centre of galaxy mass (SCOM), the star maximum-massive galaxy centre (SMGC), the star median count location (SMGL), the soft X-ray luminosity peak (GSXP), the bolometric X-ray luminosity peak (GBXP), and Compton $y$ parameter peak (GCYP). 

We first evaluate the distribution of offsets that these centroids possess, relative to the most bound particle in the cluster (MBP - defined to be the ``true'' centre).  We characterise the offsets in 2D and 3D using an analytical parameterization. Our model divides the clusters into two groups: a ``well-centred'' group described by an Exponential distribution and a ``mis-centred'' group described by Gamma distribution. The E$\Gamma$ model fits generally well to the offset distributions. The parameters for this offset model appear in Fig.~\ref{fig:cdf_bestfit}. The results are in agreement with \citet{cawthon2018dark}. Future studies of cluster density profiles could use this description to characterise mis-centred profiles.

The best-fit models in Fig.~\ref{fig:cdf_bestfit} show that the gas-based centroids GSXP, GBXP and GCYP have the smallest average offsets, as illustrated in Fig.~\ref{fig:moff_scatter}. The SITR, SMGC and SMGL centroids are clearly divided into a well-centred group and mis-centred group. About 60\% of these centroids are well-centred, consistent with the findings of \citet{cawthon2018dark} and \citet{cibirka2017codex}. However, our analysis differs in that \citet{cawthon2018dark} do not use the MBP as the true centre and have a much smaller cluster sample size, while \citet{cibirka2017codex} use only the Rayleigh distribution to describe the mis-centred group. SCOM does not have a distinct well-centred group, however, the scatter of the mis-centred group is smaller than that of the mis-centred SITR, SMGC and SMGL groups. Based on this, we conclude that SITR is the best galaxy-based centroid, in agreement with \citet{10.1111/j.1365-2966.2011.19217.x}. There is no discernible relation between centroid offset and cluster mass for the galaxy-based centroids (Fig.~\ref{fig:3d_mean}), however, the gas-based centroid offsets generally decrease with increasing cluster mass until $\sim 10^{14.4}M_{\mathrm{\odot}}$, above which the offsets increase for the largest mass clusters (possibly due to recent major mergers).

We study stacked cluster density profiles using i) all the particles, and ii) only the gas particles, in both 2D and 3D, centring each on the MBP and then on 7 observationally motivated centroids. We stack and average the density profiles in each of 20 mass bins and fit them with empirical models. The reduced $\chi^2$ values are generally greater than one, indicating that stacked profiles are not especially well fit by single-cluster models. This is improved when we measure and fit density profiles individually.  The mass bias inferred from fitting NFW profiles to the all-particle profiles typically under-estimate the true mass by $\sim$5\% for most centroids.  But the 3D profiles centred at SMGC and SMGL under-estimate cluster mass by $\sim$10\%, and thus are even more biased. For the gas density profiles, the reduced $\chi^2$ values do not depend significantly on centroid choice while the cluster masses are again generally underestimated by about $5\%$, with the SMGC and SMGL centroids again performing more poorly.

Future observational studies of galaxy cluster masses and shapes should account for the fact that cluster mis-centring can bias the inference of cluster mass from stacked data. Particularly galaxy-based centroid definitions have a significant mis-centred population whose offsets from the true centre (the MBP) are larger than the gas-based centroid definitions. In any case, we present a model for the mis-centring PDF which will allow future researchers to convolve theoretical profile models with this PDF to account for the effects of mis-centring. {It may also be fruitful to use a combination of the different centroid definitions to determine the position of the true centre (MBP) more accurately (beyond the scope of this current paper).}

\section*{Acknowledgements}
This work is financially supported by University of British Columbia, Canada's NSERC, and CIFAR. AJM and TT have received funding from the European Union's Horizon 2020 research and innovation programme under the Marie Sk\l{}odowska-Curie grant agreements No. 702971 (AJM) and No. 797794 (TT).

\medskip

\noindent\textit{Author Contributions:} N. Raza and Z. Yan contributed equally to this work.

\appendix
{
\section{Other Offset PDF Models}
\label{sec:app}
In general, the PDF of centroid offset can be written as

\begin{equation}
\begin{aligned} \mathrm{P_M}\,(r | f, \sigma, \tau) = f \times \mathrm{P_{1}}(r | \sigma)+(1-f) \times \mathrm{P_{2}}(r | \tau) ~,
\end{aligned}
\label{eq:pdf_model}
\end{equation}
where $\mathrm{P_{1}}(r | \sigma)$ describes the centred population and $\mathrm{P_{2}}(r | \tau)$ describes the mis-centred population. As shown in Table~\ref{table:pdf_models}, we explore 5 additional combinations of Gamma, Rayleigh, Gaussian and exponential distributions for $\mathrm{P_{1}}$ and $\mathrm{P_{2}}$ in addition to the E$\Gamma$ model used in the main text. $\sigma$ and $\tau$ are the width parameters of the two distributions, and $f \in [0,1]$ is the fraction of clusters that belong to the centred population. 

\begin{table}
\caption{The models we use to fit the offset distribution. In the abbreviations, 'E' stands for the exponential distribution; '$\Gamma$' for the Gamma distribution with shape parameter $k=2$; 'R' for the Rayleigh distribution, and 'G' for the Gaussian distribution.}
\centering
\begin{tabular}{lll}
\hline\hline
Name & $P_{\mathrm{1}}(r|\sigma)$  & $P_{\mathrm{2}}(r|\tau)$                                                  \\
\hline
$\Gamma\Gamma$ & $\frac{r}{\sigma^2}\exp\left(-\frac{r}{\sigma}\right)$                       & $\frac{r}{\tau^2}\exp\left(-\frac{r}{\tau}\right)$                          \\
\hline
GG             & $\frac{4\pi r^2}{\sigma^3(2\pi)^{3/2}}\exp\left(-\frac{r^2}{2\sigma^2}\right)$ & $\frac{4\pi r^2}{\tau^3(2\pi)^{3/2}}r^2\exp\left(-\frac{r^2}{2\tau^2}\right)$ \\
\hline
RR             & $\frac{r}{\sigma^2}\exp\left(-\frac{r^2}{2\sigma^2}\right)$                    & $\frac{r}{\sigma^2}\exp\left(-\frac{r^2}{2\tau^2}\right)$                     \\
\hline
E$\Gamma$      & $\frac{1}{\sigma}\exp\left(-\frac{r}{\sigma}\right)$                        & $\frac{r}{\tau^2}\exp\left(-\frac{r}{\tau}\right)$                          \\
\hline
EG             & $\frac{1}{\sigma}\exp\left(-\frac{r}{\sigma}\right)$                                        & $\frac{4\pi r^2}{\tau^3(2\pi)^{3/2}}r^2\exp\left(-\frac{r^2}{2\tau^2}\right)$ \\
\hline
ER             & $\frac{1}{\sigma}\exp\left(-\frac{r}{\sigma}\right)$                                        & $\frac{r}{\sigma^2}\exp\left(-\frac{r^2}{2\tau^2}\right)$         \\
\hline\hline
\end{tabular}
\label{table:pdf_models}
\end{table}

The fitting procedure is exactly the same as \ref{subsec:model_offset}. To compare the goodness-of-fit for the different models, we calculate the reduced $\chi^2$,
\begin{widetext}
\[
    \frac{\chi^2}{\mathrm{d.o.f}} = \frac{1}{N_{\rm bin}-3}\,\sum_{i,j} \left[{\rm C_M}(r_{{\rm off},i}|f, \sigma, \tau)-{\rm C_D}(r_{{\rm off},i})\right]^T \mathrm{Cov}^{-1}(r_{{\rm off},i}, r_{\mathrm{off},j}) \left[{\rm C_M}(r_{{\rm off},j}|f, \sigma, \tau)-{\rm C_D}(r_{{\rm off},j})\right] ~ \,
\]
\end{widetext}
\noindent where ${\rm C_D}$ is the measured offset distribution, ${\rm C_M}$ is the fitted CDF model and $\mathrm{Cov}$ is the covariance matrix of ${\rm C_D}$. The number of bins is $N_{\rm bin}=20$ while the number of degrees of freedom is $N_{\rm bin}-3 = 17$. The resulting $\chi^2$ values are listed in Tables \ref{table:chi2_3D} and \ref{table:chi2_2D}.

\begin{table}
\caption{Reduced $\chi^2$ values, in 3D, for each PDF model defined in Table~\ref{table:pdf_models} and centroid defined in Table~\ref{table:centroid_def}.}
\centering\begin{tabular}{p{10pt}p{15pt}p{15pt}p{15pt}p{15pt}p{15pt}p{15pt}p{15pt}}
\hline
\hline
 & SITR & SMGC & SMGL & SCOM & GSXP & GBXP & GCYP  \\ \hline
$\Gamma\Gamma$ & 1.07 & 8.17 & 12.11 & 5.49 & 3.2 & 2.92 & 4.1 \\ \hline
E$\Gamma$ & 1.09 & 6.82 & 10.8 & 5.33 & 1.05 & 0.9 & 0.72 \\ \hline
RR & 5.46 & 21.18 & 18.72 & 100.58 & 18.37 & 0.46 & 10.31 \\ \hline
ER & 6.04 & 19.8 & 16.66 & 98.19 & 0.57 & 0.47 & 0.73 \\ \hline
GG & 11.77 & 10.74 & 16.45 & 2.86 & 2.64 & 4.09 & 1.14 \\ \hline
EG & 13.74 & 10.93 & 16.2 & 2.82 & 2.85 & 3.96 & 1.22 \\ \hline
\hline
\end{tabular}
\label{table:chi2_3D}
\end{table}

\begin{table}
\caption{Reduced $\chi^2$ values, in 2D, for each PDF model defined in Table~\ref{table:pdf_models} and centroid defined in Table~\ref{table:centroid_def}.}
\centering\begin{tabular}{p{10pt}p{15pt}p{15pt}p{15pt}p{15pt}p{15pt}p{15pt}p{15pt}}
\hline
\hline
 & SITR & SMGC & SMGL & SCOM & GSXP & GBXP & GCYP  \\ \hline
$\Gamma\Gamma$ & 1.36 & 2.43 & 6.82 & 1.27 & 4.13 & 4.01 & 3.16 \\ \hline
E$\Gamma$ & 1.2 & 2.56 & 6.89 & 1.1 & 1.58 & 3.18 & 2.24 \\ \hline
RR & 2.44 & 8.37 & 4.64 & 39.34 & 16.12 & 1.04 & 0.87 \\ \hline
ER & 2.88 & 9.54 & 5.68 & 40.15 & 0.96 & 1.08 & 0.97 \\ \hline
GG & 14.91 & 13.41 & 19.66 & 2.45 & 3.16 & 7.27 & 3.73 \\ \hline
EG & 15.12 & 13.72 & 20.97 & 2.34 & 3.14 & 7.38 & 3.81 \\ \hline
\hline
\end{tabular}
\label{table:chi2_2D}
\end{table}

Surveying across all the values from both tables, we see that the E$\Gamma$ model (see Table \ref{table:pdf_models}) generally provides the best description of the various centroid distributions, since it consistently produces one of the lowest (if not \textit{the} lowest) reduced $\chi^2$ values for all centroid definitions.

\begin{landscape}
\begin{figure}
	\includegraphics[width=0.8\paperheight] {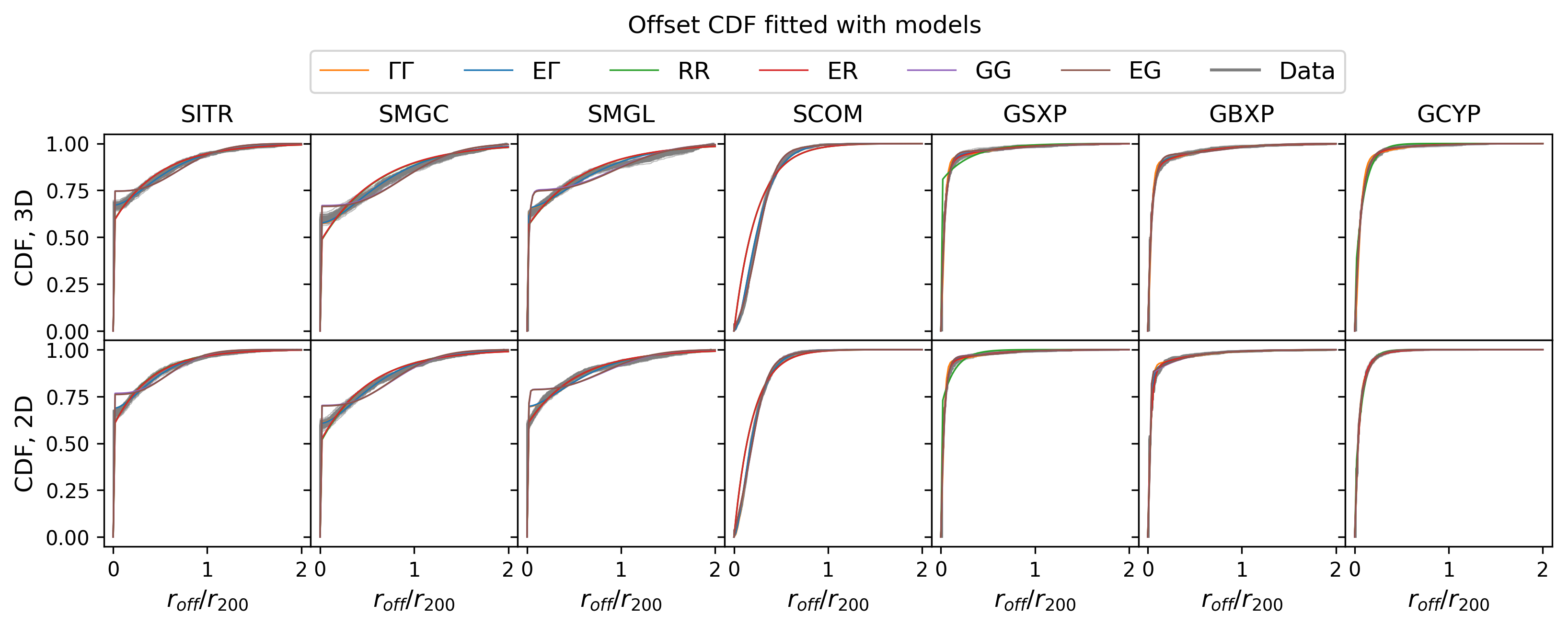}
    \caption{The measured (grey) and 6 fitted models (as defined in Table \ref{table:pdf_models}) of offset CDF from the MBP, for all seven centroids analysed. The upper panels show the 3D analysis and lower panels 2D. The grey collection of lines are the results of all bootstrap re-sampled offset CDFs.}
    \label{fig:cdf_fittings}
\end{figure}
 \end{landscape}
}


\bibliographystyle{mnras}
\bibliography{main}


\label{lastpage}
\end{document}